\documentclass[conference]{IEEEtran}
\usepackage{amsmath}
\usepackage{amsfonts}
\usepackage{cite}
\usepackage{amssymb}
\usepackage{graphicx}
\usepackage[usenames,dvipsnames]{color}
\usepackage{epstopdf}
\usepackage[all]{xy}
\usepackage[]{mcode}
\usepackage{listings}

\usepackage{subcaption}
\usepackage{longtable}

\begin{document}

\title{Power Allocation for Adaptive OFDM Index Modulation in Cooperative Networks}

\author{\IEEEauthorblockN{Shuping Dang, Gaojie Chen and Justin P. Coon}
\IEEEauthorblockA{Department of Engineering Science\\University of Oxford\\Parks Road, Oxford, OX1 3PJ, UK\\
Email: \{shuping.dang, gaojie.chen, justin.coon\}@eng.ox.ac.uk}}

\maketitle

\begin{abstract}
In this paper, we propose a power allocation strategy for the adaptive orthogonal frequency-division multiplexing (OFDM) index modulation (IM) in cooperative networks. The allocation strategy is based on the Karush-Kuhn-Tucker (KKT) conditions, and aims at maximizing the average network capacity according to the instantaneous channel state information (CSI). As the transmit power at source and relay is constrained separately, we can thus formulate an optimization problem by allocating power to active subcarriers. Compared to the conventional uniform power allocation strategy, the proposed dynamic strategy can lead to a higher average network capacity, especially in the low signal-to-noise ratio (SNR) region. The analysis is also verified by numerical results produced by Monte Carlo simulations. By applying the proposed power allocation strategy, the efficiency of adaptive OFDM IM can be enhanced in practice, which paves the way for its implementation in the future, especially for cell-edge communications.

\end{abstract}


\IEEEpeerreviewmaketitle

\section{Introduction}
Stemming from parallel combinatorial signaling, spatial modulation (SM) and orthogonal frequency-division multiplexing (OFDM) index modulation (IM) have been regarded as two of the most promising modulation techniques for next generation networks \cite{764929,6678765,7469311,7509396,7879247}. SM is designed for multiple-input-multiple-output (MIMO) systems and exploits the spatial dimension, while OFDM IM is mainly applied to multicarrier systems by conveying information via the frequency dimension. The investigations into SM are relatively full-fledged and a complete framework regarding performance analysis, resource allocation and practical implementation has been constructed \cite{6678765}. Compared to SM, OFDM IM is relatively new and many aspects require attention. The concept of OFDM IM is first proposed in \cite{6587554}, and then systematically improved and generalized in \cite{7112187} and \cite{7234862}, respectively. Subsequently, the transmission rate of OFDM IM is analyzed in \cite{7330022} and the extension of OFDM IM to cooperative networks is presented in \cite{zhunbeijiao}, in which the outage performance, network capacity and error performance of OFDM IM are analyzed in detail. 

However, in most existing works, the transmit power is simply allocated to each subcarrier in a uniform manner, and the analysis of the power allocation problem is still lacking. As shown in conventional OFDM systems and SM systems, an optimized power allocation strategy would significantly enhance the system performance in terms of throughput and reliability \cite{1296633,7809043,7812789}. In \cite{5992826} and \cite{4533749}, power allocation problems are analyzed and solved for OFDM systems with DF relays based on the Karush-Kuhn-Tucker (KKT) conditions and KKT multipliers. A similar approach is also proved to be effective for OFDM-based cognitive radio systems and multiuser scenarios \cite{4411682,1437359}. Following the analysis given in \cite{zhunbeijiao}, we thereby adopt a similar KKT conditions-based approach and carry out the analysis regarding power allocation for adaptive OFDM IM in cooperative networks in this paper. Finally, we formulate the optimization problem and propose a dynamic power allocation strategy aiming at maximizing the average network capacity. Numerical results verify that with the proposed power allocation strategy, the average network capacity can be improved compared to that with uniform power allocation strategy, especially in the low signal-to-noise ratio (SNR) region. This makes the proposed strategy particularly useful for cell-edge communications.

The rest of the paper is organized as follows. Section \ref{sm} presents the system model. We then formulate the optimization problem and analyze the power allocation strategy in Section \ref{pfa}. Subsequently, numerical results are shown and discussed in Section \ref{nr}. Finally, the paper is concluded in Section \ref{c}.

\section{System Model}\label{sm}
\begin{figure}[!t]
\centering
\includegraphics[width=3.5in]{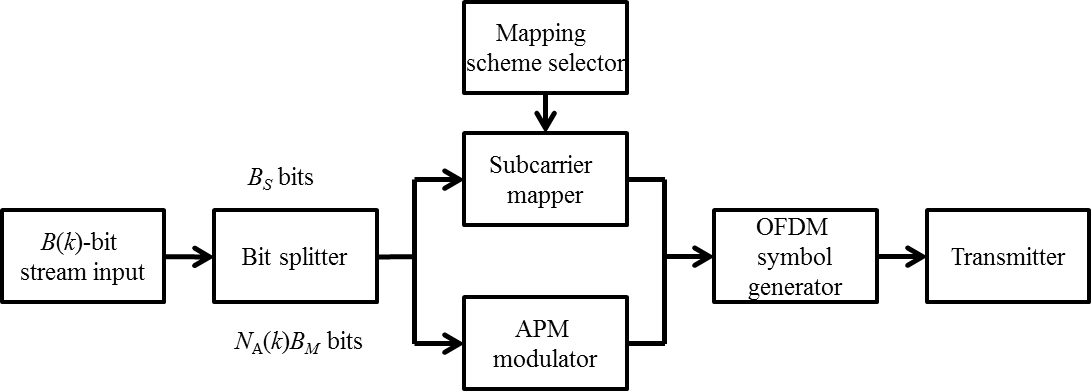}
\caption{System block diagram of adaptive OFDM IM, reproduced from \cite{zhunbeijiao} with permission.}
\label{sys}
\end{figure}
\subsection{System framework}
Following the system model constructed in \cite{zhunbeijiao}, we consider a two-hop OFDM IM system with $N_T$ subcarriers, in which one source, one relay and one destination exist. The transmit power at source and relay is constrained separately by the same bound $P_t$. This system operates in a slow frequency-selective fading environment and the fading on each subcarrier can be regarded as independent and identically distributed (i.i.d.). Then, $N_S$ ($1\leq N_S <N_T$) subcarriers are selected by a certain criterion (details given in Section \ref{mappingshemesectchap}) to construct a \textit{mapping scheme} for OFDM IM\footnote{Note that, by such a selection process, the constructed mapping scheme is NOT the same as the inherent activation pattern used in OFDM IM systems.}. After that, a bit stream with a variable-length $B(k)=B_S+N_A(k)B_M$ can be modulated by both the subcarrier activation pattern and the conventional $M$-ary amplitude and phase modulation (APM) scheme (e.g. $M$-PSK and $M$-QAM), where $k\in\{1,2,3,\dots,2^{N_S}\}$ denoting the index of the subcarrier activation pattern and $N_A(k)$ is the number of active subcarriers for the $k$th pattern; $B_S=N_S$ is the length of the bit stream which will be modulated by the subcarrier activation pattern in an on-off keying (OOK) manner, and on each active subcarrier, a symbol generated by a $B_M$-bit stream is transmitted. This is termed adaptive OFDM IM and the system block diagram is illustrated in Fig. \ref{sys}. However, one might note that, as the subcarriers are activated in an OOK manner, there is a possibility that all subcarriers are inactive when an all-zero $B_S$-bit stream is transmitted (we denote this case as $k=1$). This is termed \textit{zero-active subcarrier dilemma}, and a \textit{complementary subcarrier} from those $N_T-N_S$ unselected subcarriers will be activated to undertake the transmission of at least one APM symbol\footnote{More details of this dilemma can be found at \cite{zhunbeijiao}.}.

Meanwhile, we assume there is no direct transmission link between source and destination due to deep fading and the signal propagation must go through relay. Also, a half-duplex decode-and-forward (DF) forwarding protocol is adopted at the relay, and two orthogonal temporal phases are required for one complete transmission from source to relay, and from relay to destination.

\subsection{Channel model}
It is assumed that the channels in the first and second hops are slow frequency-selective Rayleigh faded with exponentially distributed channel gains. Here, the slow property indicates that quasi-static block fading channels are considered and the channel gains are random but would remain unchanged for a sufficiently large period of time \cite{7471479}, so that the overheads for transmitting the selected mapping scheme and information of power allocation via feedforward links to relay and destination for decoding purposes are negligible \cite{5752793}. Denoting the set of all subcarriers as $\mathcal{N}=\{1,2,\dots,N_T\}$,  $\forall~n\in\mathcal{N}$, the probability density function (PDF) and the cumulative distribution function (CDF) of the channel gain $|h_i(n)|^2$ are
\begin{equation}\label{channelpdfcdf}
f_{i}(s)=\mathrm{exp}\left(-s/\mu_i\right)/\mu_i~~\Leftrightarrow~~F_{i}(s)=1-\mathrm{exp}\left(-s/\mu_i\right)
\end{equation}
where $\mu_i$ denotes the average channel gain of the $i$th hop.

\subsection{Mapping scheme selections}\label{mappingshemesectchap}
In OFDM IM systems, a general $N_T \times 1$ transmit OFDM block in frequency domain can be written as
\begin{equation}
\mathbf{x}=[x(1),x(2),\dots,x(N_T)]^T\in\mathbb{C}^{N_T\times 1},
\end{equation}
where $(\cdot)^T$ denotes the matrix transpose operation and $\mathbb{C}$ is the field of complex numbers.

After selecting an arbitrary $c$th mapping scheme, $c\in\mathcal{C}$, where $\mathcal{C}$ is the set of all possible mapping schemes, and obtaining a subset $\mathcal{N}_S(c)\subset\mathcal{N}$ for OFDM IM, the reduced OFDM block determined by the $B(k)$-bit stream is given by
\begin{equation}\label{545msmdx2}
\mathbf{x}(k)=[x(m_1,1),x(m_2,2),\dots,x(m_{N_S},N_S)]^T\in\mathbb{C}^{N_S\times 1},
\end{equation}
where
\begin{equation}
x(m_n,n)=\begin{cases}
\chi_{m_n}, ~~~~n\in\mathcal{N}_A(k)\\
0,~~~~~~~~\mathrm{otherwise}
\end{cases}
\end{equation}
corresponds to the data symbol transmitted on the $n$th subcarrier, and $\mathcal{N}_A(k)\subseteq\mathcal{N}_S(c)$ is the subset of $N_A(k)$ active subcarriers for the $k$th subcarrier activation pattern; $\chi_{m_n}$ is the $M$-ary APM symbol and we can normalize it by $\chi_{m_n}\chi_{m_n}^*=1$.

Therefore, for $k\neq 1$ (i.e. there is at least one active subcarrier), we can obtain the received SNR in the $i$th hop for the $n$th subcarrier when the $k$th subcarrier activation pattern is utilized by
\begin{equation}\label{receivedsnr}
\gamma_i(k,n)=\begin{cases}\frac{P_{t,i}(k,n)}{N_0}|h_i(c,n)|^2,~~~~~~\forall~n\in\mathcal{N}_A(k)\\
0,~~~~~~~~~~~~~~~~~~~~~~~~~~~\forall~n\in\mathcal{N}_S(c)\setminus \mathcal{N}_A(k)
\end{cases}
\end{equation}
where $P_{t,i}(k,n)$ is the allocated transmit power to the $n$th subcarrier in the $i$th hop and $N_0$ is the additive white Gaussian noise (AWGN) power; $|h_i(c,n)|^2$ is the channel gain regarding the $n$th subcarrier after performing mapping scheme selection.

On the other hand, when $k=1$, we can have
\begin{equation}
\gamma_i(1,\tilde{n}_i)=\frac{P_t}{N_0}|h_i(\tilde{n}_i)|^2~~\mathrm{and}~~\gamma_i(1,n)=0,~\forall~n\in\mathcal{N}_S(c),
\end{equation}
where $\tilde{n}_i$ denotes the index of the complementary subcarrier selected for the $i$th hop.

\subsubsection{Decentralized mapping scheme selection}
Now, we can specify the method of mapping scheme selection. Two mapping scheme selection methods are introduced in this paper, which are applied to different types of networks depending on the processing capability of the relay node. If both source and relay can get access to CSI and perform mapping scheme selections independently, we can adopt the decentralized mapping scheme selection method in each hop by the criterion:
\begin{equation}\label{mappselect1}
\hat{c}_i=\underset{c\in\mathcal{C}}{\arg\max}\left\lbrace\sum_{n\in\mathcal{N}_S(c)}\gamma_i(2^{N_S},n)\right\rbrace.
\end{equation}
Also, the complementary subcarrier in each hop can be selected by
\begin{equation}\label{xuanzejizhibuchong}
\tilde{n}_i=\underset{n\in\mathcal{N}\setminus\mathcal{N}_S(\hat{c}_i)}{\arg\max}|h_i(n)|^2.
\end{equation}

\subsubsection{Centralized mapping scheme selection}
On the other hand, for a simple relay which is unable to perform mapping scheme selection due to limited system complexity and processing capability, we can utilize the centralized mapping scheme selection method, by which the mapping scheme selection is only performed at the source and utilized by the relay. The selection criterion can be written as
\begin{equation}\label{mappselect3}
\begin{split}
&\hat{c}_1=\hat{c}_2=\hat{c}\\
&=\underset{c\in\mathcal{C}}{\arg\max}\left\lbrace\sum_{n\in\mathcal{N}_S(c)}\min\left\lbrace\gamma_1(2^{N_S},n),\gamma_2(2^{N_S},n)\right\rbrace\right\rbrace,
\end{split}
\end{equation}
and the complementary subcarrier can be similarly selected by
\begin{equation}\small
\tilde{n}_1=\tilde{n}_2=\tilde{n}=\underset{n\in\mathcal{N}\setminus\mathcal{N}_S(\hat{c})}{\arg\max}\min\left\lbrace|h_1(n)|^2,|h_2(n)|^2\right\rbrace.
\end{equation}

\begin{figure}[!t]
\centering
\includegraphics[width=3.5in]{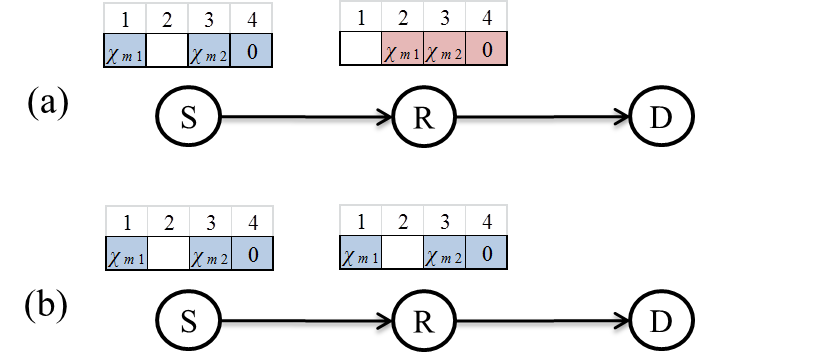}
\caption{An example of (a): a two-hop system with the decentralized mapping scheme selection; (b): a two-hop system with the centralized mapping scheme selection, when $N_T=4$, $N_S=3$ and $N_A(k)=2$.}
\label{mapping}
\end{figure}

The two-hop systems with decentralized and centralized mapping scheme selections are illustrated in Fig. \ref{mapping} for clarity.

\subsection{Network capacity}
By the max-flow min-cut theorem \cite{5071270}, the \textit{average network capacity} in two-hop networks can be expressed by
\begin{equation}\label{capacityexressectsasd235}
\begin{split}
&\bar{C}={\mathbb{E}}\left\lbrace C(k)\right\rbrace~~[\mathrm{bit/s/Hz}],
\end{split}
\end{equation}
where $\mathbb{E}\{\cdot\}$ represents the expectation over all channels and subcarrier activation patterns;   $C(k)$ is the network capacity when the $k$th subcarrier activation pattern is utilized and can be written as
\begin{equation}\label{dsadoutsctutasn}\small
C(k)=\begin{cases}
\frac{1}{2}\min\left\lbrace\log_2\left(1+\gamma_1(1,\tilde{n}_1)\right),\log_2\left(1+\gamma_2(1,\tilde{n}_2)\right)\right\rbrace,\\~~~~~~~~~~~~~~~~~~~~~~~~~~~~~~~~~~~~~~~~~~~~~~~~~~~~~~~~\mathrm{if} ~k=1\\
\underset{n\in\mathcal{N}_A(k)}{\sum}\frac{1}{2}\min\left\lbrace\log_2\left(1+\gamma_1(k,{n})\right),\log_2\left(1+\gamma_2(k,{n})\right)\right\rbrace,\\~~~~~~~~~~~~~~~~~~~~~~~~~~~~~~~~~~~~~~~~~~~~~~~~~~~~~~~~\mathrm{if} ~k>1\\
\end{cases}
\end{equation}
The average network capacity will be adopted in the following analysis to evaluate the system throughput, and the maximization of the average network capacity is the objective of the formulated optimization problem in the next section\footnote{For a typical OFDM system, there are two kinds of optimization problems: 1) maximizing the network capacity under a limited total transmit power; 2) minimizing the total transmit power beyond a threshold network capacity \cite{liu2013resource}. Because the user experience is regarded as one of the most important metrics in next generation networks\cite{6736752}, we take the former optimization scenario in this paper.}.

\section{Problem Formulation and Analysis}\label{pfa}
\subsection{Dynamic power allocation strategy}
For the adaptive OFDM IM systems in two-hop networks introduced in the previous section, we can formulate the optimization problem infra:
\begin{equation}\label{problemformulationeq}
\begin{split}
&~~~~~~\max_{\begin{subarray}{c}\mathbf{P}_{t,1}(k)\\\mathbf{P}_{t,2}(k)\end{subarray}}\left\lbrace\bar{C}\right\rbrace\\
&\mathrm{s.t.}\sum_{n\in\mathcal{N}_A(k)}P_{t,1}(k,n)\leq P_{t},~\sum_{n\in\mathcal{N}_A(k)}P_{t,2}(k,n)\leq P_{t}\\
&~~~~P_{t,1}(k,n)\geq 0,~P_{t,2}(k,n)\geq 0,~\forall k\in\mathcal{K},n\in\mathcal{N}_A(k).
\end{split}
\end{equation}
where $\mathbf{P}_{t,i}(k)=\left[P_{t,i}(k,n_1),\dots,P_{t,i}(k,n_{N_A(k)})\right]^T$, $\forall~k\in\mathcal{K}$. By (\ref{capacityexressectsasd235}), we can reduce the expression of $\bar{C}$ to
\begin{equation}\label{reducedcapacityexressectsasd235}
\begin{split}
&\bar{C}=\underset{k\in\mathcal{K}}{\mathbb{E}}\left\lbrace\underset{h_{1}(n),h_{2}(n)}{\mathbb{E}} \left\lbrace C(k)\right\rbrace\right\rbrace~~[\mathrm{bit/s/Hz}],
\end{split}
\end{equation}
From (\ref{reducedcapacityexressectsasd235}), we observe that as long as the network capacity of each instant can be optimized, the average network capacity will be optimized. As a result, we can equivalently transfer the optimization problem formulated in (\ref{problemformulationeq}) to
\begin{equation}\label{problemformdsjakdjh2ulationeq}
\begin{split}
&~~~~~~\max_{\begin{subarray}{c}\mathbf{P}_{t,1}(k)\\\mathbf{P}_{t,2}(k)\end{subarray}}\left\lbrace {C}(k)\right\rbrace\\
&\mathrm{s.t.}\sum_{n\in\mathcal{N}_A(k)}P_{t,1}(k,n)\leq P_{t},~\sum_{n\in\mathcal{N}_A(k)}P_{t,2}(k,n)\leq P_{t}\\
&~~~~P_{t,1}(k,n)\geq 0,~P_{t,2}(k,n)\geq 0,~\forall k\in\mathcal{K},n\in\mathcal{N}_A(k).
\end{split}
\end{equation}

\subsubsection{Decentralized mapping scheme selection}
When $k=1$, the complementary subcarrier $\tilde{n}_i$ is the only one active subcarrier in the $i$th hop and thus all transmit power $P_{t}$ will be allocated to it. For such a special case, there is only a unique solution to (\ref{problemformdsjakdjh2ulationeq}), and the power allocation is easy to deal with. On the other hand, when $k\neq 1$, because the power allocations in two hops are independent when the decentralized mapping scheme selection method is employed, we can further split the reduced optimization problem formulated in (\ref{problemformdsjakdjh2ulationeq}) into 
\begin{equation}\label{problemformdsjakdjh2ulationeq11}
\begin{split}
&~~~~~~\max_{\begin{subarray}{c}\mathbf{P}_{t,i}(k)\end{subarray}}\left\lbrace {C}_i(k)\right\rbrace\\
&\mathrm{s.t.}\sum_{n\in\mathcal{N}_A(k)}P_{t,i}(k,n)\leq P_{t},~P_{t,i}(k,n)\geq 0,\\
&~~~~~~~~\forall k\in\mathcal{K},n\in\mathcal{N}_A(k).
\end{split}
\end{equation}
for each hop, where $C_i(k)=\sum_{n\in\mathcal{N}_A(k)}\frac{1}{2}\log_2\left(1+\gamma_i(k,{n})\right)$. Besides, by (\ref{receivedsnr}), it is obvious that $C_i(k)$ is continuous and differentiable with respect to $P_{t,i}(k,n)$, $\forall~n\in\mathcal{N}_A(k)$.

We can find the optimization problem formulated in (\ref{problemformdsjakdjh2ulationeq11}) to be a standard \textit{nonlinear programming problem} \cite{patriksson2013nonlinear}. Now, by employing the KKT conditions-based approach, we can construct the KKT function for the $i$th hop as follows\footnote{For simplicity, we will use $\mathcal{L}_i$ as a shorthand in the following analysis.}
\begin{equation}\small
\begin{split}
&\mathcal{L}_i(P_{t,i}(k,n_1),P_{t,i}(k,n_2),\dots,P_{t,i}(k,n_{N_A(k)}),\\
&~~~~~~~~~~~~~~~~~~~~~~~~\epsilon_i,\varepsilon_i(n_1),\varepsilon_i(n_2),\dots,\varepsilon_i(n_{N_A(k)}))\\
&=C_i(k)+\epsilon_i\left(P_{t}-\sum_{n\in\mathcal{N}_A(k)}P_{t,i}(k,n)\right)+\sum_{n\in\mathcal{N}_A(k)}\varepsilon_i(n)P_{t,i}(k,n),
\end{split}
\end{equation}
where $\epsilon_i$ and $\varepsilon_i(n)$ are the KKT multipliers for the $i$th hop, $\forall~n\in\mathcal{N}_A(k)$.

Subsequently, we can derive the KKT conditions to be
\begin{equation}\label{partialheheseqs6523462}\small
\begin{cases}
\frac{\partial\mathcal{L}_i}{\partial P_{t,i}(k,n_1)}=\frac{|h_i(\hat{c}_i,n_1)|^2}{(2\ln 2)\left(N_0+|h_i(\hat{c}_i,n_1)|^2P_{t,i}(k,n_1)\right)}-\epsilon_i+\varepsilon_i(n_1)=0\\
\frac{\partial\mathcal{L}_i}{\partial P_{t,i}(k,n_2)}=\frac{|h_i(\hat{c}_i,n_2)|^2}{(2\ln 2)\left(N_0+|h_i(\hat{c}_i,n_2)|^2P_{t,i}(k,n_2)\right)}-\epsilon_i+\varepsilon_i(n_2)=0\\
\quad\quad\quad\quad\vdots\\
\frac{\partial\mathcal{L}_i}{\partial\epsilon_i}=P_{t}-\underset{n\in\mathcal{N}_A(k)}{\sum}P_{t,i}(k,n)=0\\
\varepsilon_i(n_1)P_{t,i}(k,n_1)=0\\
\varepsilon_i(n_2)P_{t,i}(k,n_2)=0\\
\quad\quad\quad\quad\vdots\\
\varepsilon_i(n_{N_A(k)})P_{t,i}(k,n_{N_A(k)})=0\\
\end{cases}
\end{equation}

Solving the equation set given in (\ref{partialheheseqs6523462}) yields the optimal ${\mathbf{P}}_{t,i}^*(k)$. In order to perform numerical evaluation later, we can generally express the standard waterfilling solution given by
\begin{equation}\label{ejdhaskjhd2}
{P}_{t,i}^*(k,\dot{n})=\left[\frac{1}{(2\ln 2)(\epsilon_i-\varepsilon_i(\dot{n}))}-\frac{N_0}{|h_i(\hat{c}_i,\dot{n})|^2}\right]^+,
\end{equation}
where $[x]^+=\max\left\lbrace 0,x\right\rbrace$; $\epsilon_i$ and $\varepsilon_i(\dot{n})$ satisfy
\begin{equation}\small
P_t-\sum_{n\in\mathcal{N}_A(k)}\left(\frac{1}{(2\ln 2)(\epsilon_i-\varepsilon_i({n}))}-\frac{N_0}{|h_i(\hat{c}_i,{n})|^2}\right)=0
\end{equation}
and
\begin{equation}
\varepsilon_i({n})\left[\frac{1}{(2\ln 2)(\epsilon_i-\varepsilon_i(\dot{n}))}-\frac{N_0}{|h_i(\hat{c}_i,\dot{n})|^2}\right]^+=0.
\end{equation}
A general and closed-form expression of (\ref{ejdhaskjhd2}) independent from $\epsilon_i$ and $\varepsilon_i(\dot{n})$ does not exist, since the solution is associated with the quantitative relation among all channel gains. However, for a given network realization, the solution can be analytically determined by an iterative algorithm (see \textit{Theorem 1} in \cite{5430230}), and the algorithm is implemented by the MATLAB function \texttt{fmincon}.

\subsubsection{Centralized mapping scheme selection}
However, when the centralized mapping scheme selection is performed, the power allocations in the first and second hop are not independent anymore and should be considered jointly. As a consequence, we cannot split (\ref{problemformdsjakdjh2ulationeq}) into (\ref{problemformdsjakdjh2ulationeq11}) for two hops independently. In this case, we have to integrate the channels in two hops into a \textit{link} and define the link gain as
\begin{equation}
|l(\hat{c},n)|^2=\min\left\lbrace|h_1(\hat{c},n)|^2,|h_2(\hat{c},n)|^2\right\rbrace.
\end{equation}

For $k=1$, again, there is only one possibility of power allocation and we can simply allocate all transmit power to the complementary subcarrier $\tilde{n}$. On the other hand, when $k\neq 1$, we can employ the KKT conditions-based approach and construct the KKT function for a whole end-to-end link as follows:
\begin{equation}\small
\begin{split}
&\mathcal{L}(P_{t}(k,n_1),P_{t}(k,n_2),\dots,P_{t}(k,n_{N_A(k)}),\\
&~~~~~~~~~~~~~~~~~~~~~~~~\epsilon,\varepsilon(n_1),\varepsilon(n_2),\dots,\varepsilon(n_{N_A(k)}))\\
&=C(k)+\epsilon\left(P_{t}-\sum_{n\in\mathcal{N}_A(k)}P_{t}(k,n)\right)+\sum_{n\in\mathcal{N}_A(k)}\varepsilon(n)P_{t}(k,n),
\end{split}
\end{equation}
where $\epsilon$ and $\varepsilon(n)$ are the KKT multipliers, $\forall~n\in\mathcal{N}_A(k)$.

Similar to the decentralized case, we have the KKT conditions as follows:
\begin{equation}\label{censdsa2partialheheseqs6523462}\small
\begin{cases}
\frac{\partial\mathcal{L}}{\partial P_{t}(k,n_1)}=\frac{|l(\hat{c},n_1)|^2}{(2\ln 2)\left(N_0+|l(\hat{c},n_1)|^2P_{t}(k,n_1)\right)}-\epsilon+\varepsilon(n_1)=0\\
\frac{\partial\mathcal{L}}{\partial P_{t}(k,n_2)}=\frac{|l(\hat{c},n_2)|^2}{(2\ln 2)\left(N_0+|l(\hat{c},n_2)|^2P_{t}(k,n_2)\right)}-\epsilon+\varepsilon(n_2)=0\\
\quad\quad\quad\quad\vdots\\
\frac{\partial\mathcal{L}}{\partial\epsilon}=P_{t}-\underset{n\in\mathcal{N}_A(k)}{\sum}P_{t}(k,n)=0\\
\varepsilon(n_1)P_{t}(k,n_1)=0\\
\varepsilon(n_2)P_{t}(k,n_2)=0\\
\quad\quad\quad\quad\vdots\\
\varepsilon(n_{N_A(k)})P_{t}(k,n_{N_A(k)})=0\\
\end{cases}
\end{equation}

Solving the equation set given in (\ref{censdsa2partialheheseqs6523462}) yields the optimal ${\mathbf{P}}_{t,1}^*(k)={\mathbf{P}}_{t,2}^*(k)=\left[{P}_t^*(k,n_1),{P}_t^*(k,n_2),\dots,{P}_t^*(k,n_{N_A(k)})\right]^T$. Again, we can only generally express the solution of optimal power allocated to each active subcarrier by
\begin{equation}\label{ejdhaskjhdheskjdskj22}
{P}_{t,1}^*(k,\dot{n})={P}_{t,2}^*(k,\dot{n})=\left[\frac{1}{(2\ln 2)(\epsilon-\varepsilon(\dot{n}))}-\frac{N_0}{|l(\hat{c},\dot{n})|^2}\right]^+,
\end{equation}
where $\epsilon$ and $\varepsilon(\dot{n})$ satisfy
\begin{equation}\small
P_t-\sum_{n\in\mathcal{N}_A(k)}\left(\frac{1}{(2\ln 2)(\epsilon-\varepsilon({n}))}-\frac{N_0}{|l(\hat{c},{n})|^2}\right)=0
\end{equation}
and
\begin{equation}
\varepsilon({n})\left[\frac{1}{(2\ln 2)(\epsilon-\varepsilon({n}))}-\frac{N_0}{|l(\hat{c},{n})|^2}\right]^+=0.
\end{equation}
There does not exist a general and closed-form expression of (\ref{ejdhaskjhdheskjdskj22}) either, and for a given network realization, the solution can be determined by the iterative algorithm presented in \cite{5430230}, which is implemented by the MATLAB function \texttt{fmincon}.

\subsection{Uniform power allocation strategy}
In contrast to our proposed dynamic power allocation strategy depending on CSI, the conventional uniform power allocation strategy is simple and does not rely on CSI, but will lead to a suboptimal average capacity. We also briefly introduce the uniform power allocation strategy here, as it is a commonly used strategy and will be adopted as the comparison benchmark in next section. When adopting the uniform power allocation strategy, the transmit power allocated to each subcarrier is the same and can be written as
\begin{equation}
{P}_{t,1}(k,\dot{n})={P}_{t,2}(k,\dot{n})=\frac{P_t}{N_A(k)}.
\end{equation}

\begin{figure*}[t!]
    \centering
    \begin{subfigure}[t]{0.5\textwidth}
        \centering
        \includegraphics[width=3.5in]{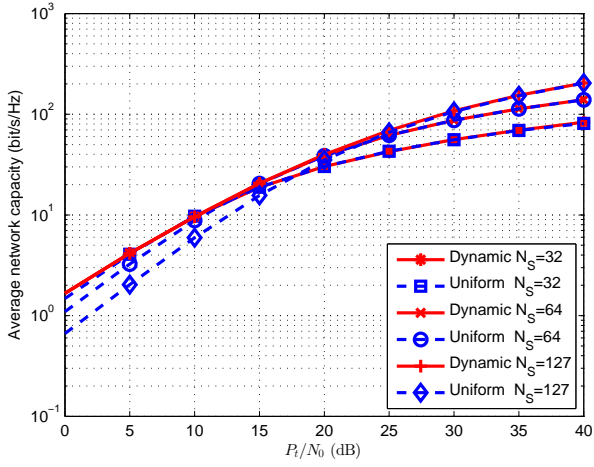}
        \caption{Decentralized mapping scheme selection: $N_T=128$.}
    \end{subfigure}%
~
    \begin{subfigure}[t]{0.5\textwidth}
        \centering
        \includegraphics[width=3.5in]{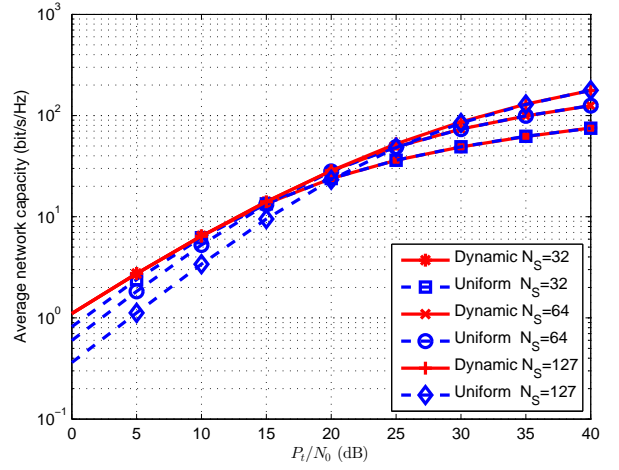}
        \caption{Centralized mapping scheme selection: $N_T=128$.}
    \end{subfigure}
~
    \centering
    \begin{subfigure}[t]{0.5\textwidth}
        \centering
        \includegraphics[width=3.5in]{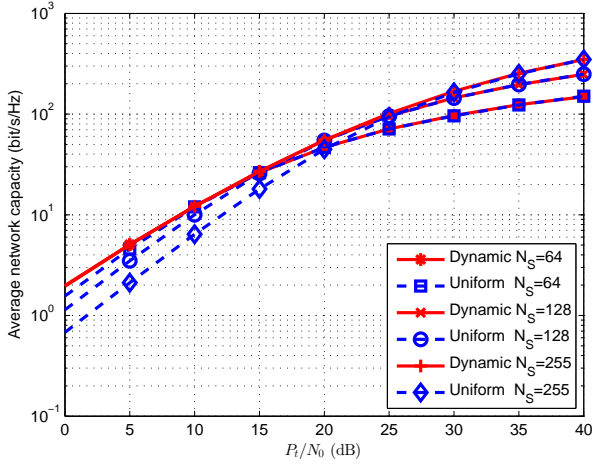}
        \caption{Decentralized mapping scheme selection: $N_T=256$.}
    \end{subfigure}%
~
    \begin{subfigure}[t]{0.5\textwidth}
        \centering
        \includegraphics[width=3.5in]{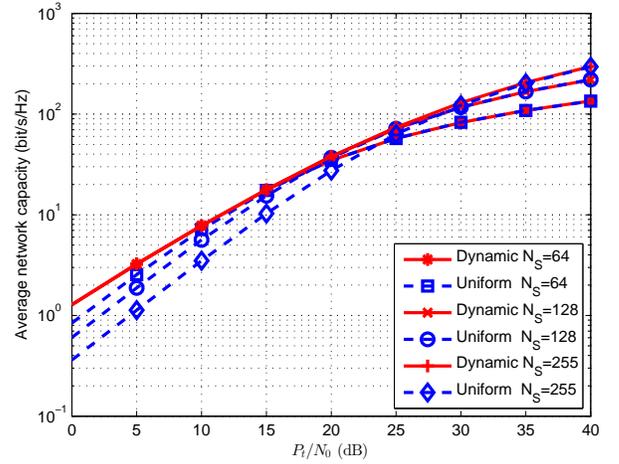}
        \caption{Centralized mapping scheme selection: $N_T=256$.}
    \end{subfigure}
    \caption{Average network capacity vs. ratio of total transmit power to noise power $P_t/N_0$.}
    \label{fig_capacity_vs_pt}
\end{figure*}

\section{Numerical Results}\label{nr}
We normalize $\mu_1=1$, $\mu_2=1$ and $N_0=1$ and carried out the numerical simulations regarding the relation between $P_t/N_0$ and the average network capacity for systems adopting different mapping scheme selection methods and power allocation strategies. In this paper, we directly employ the MATLAB function \texttt{fmincon} to solve the optimal ${\mathbf{P}}_{t,1}^*(k)$ and ${\mathbf{P}}_{t,2}^*(k)$. The numerical results are shown in Fig. \ref{fig_capacity_vs_pt}. From this figure, we can observe that the systems with the dynamic power allocation strategy outperform the systems with the uniform power allocation strategy in terms of average network capacity, which verifies the efficiency of our proposed dynamic strategy for the adaptive OFDM IM in two-hop networks. 

Besides, it is well known that the advantage brought by the dynamic allocation strategy will diminish when $P_t/N_0$ increases. This is because when a sufficiently large total transmit power can be provided, the impacts of channel/link gains on the capacity are relatively trivial, and the power allocated to each subcarrier will converge to $P_t/N_A(k)$ (i.e. the uniform strategy). Mathematically, by (\ref{ejdhaskjhd2}) and (\ref{ejdhaskjhdheskjdskj22}), this property can be derived by
\begin{equation}\label{jixiantuidao1}\small
\begin{split}
&\lim_{P_t\rightarrow\infty}{P}_{t,i}^*(k,\dot{n})\\
&=\lim_{P_t\rightarrow\infty}\left(\frac{P_t+\sum_{n\in\mathcal{N}_A(k)}\frac{N_0}{|h_i(\hat{c}_i,{n})|^2}}{N_A(k)}-\frac{N_0}{|h_i(\hat{c}_i,\dot{n})|^2}\right)=\frac{P_t}{N_A(k)},
\end{split}
\end{equation}
and 
\begin{equation}\label{jixiantuidao2}\small
\begin{split}
&\lim_{P_t\rightarrow\infty}{P}_{t,1}^*(k,\dot{n})=\lim_{P_t\rightarrow\infty}{P}_{t,2}^*(k,\dot{n})\\
&=\lim_{P_t\rightarrow\infty}\left(\frac{P_t+\sum_{n\in\mathcal{N}_A(k)}\frac{N_0}{|l(\hat{c},{n})|^2}}{N_A(k)}-\frac{N_0}{|l(\hat{c},\dot{n})|^2}\right)=\frac{P_t}{N_A(k)},
\end{split}
\end{equation}
for decentralized and centralized cases, respectively. This indicates that our proposed dynamic power allocation strategy will play a more important role and result in a significant gain of channel capacity for  cell-edge communications. 

Another phenomenon shown in the numerical results is that systems with different number of selected subcarriers $N_S$ have a similar average network capacity in the low SNR region, when the dynamic power allocation strategy is utilized. This can be explained by the fact that all transmit power will be allocated to the strongest channel/link at low SNR, regardless $N_S$. In this scenario, the OFDM IM system will degrade to a conventional OFDM system unable to convey information by the subcarrier activation pattern. As a consequence, the gain of average network capacity brought by the dynamic power allocation strategy can also be viewed as yielded by a mode selection mechanism switching between the OFDM IM mode and the conventional OFDM mode.

\section{Conclusion}\label{c}
We proposed a power allocation strategy for the adaptive OFDM IM in cooperative networks. Under a constrained total transmit power, the power allocation strategy utilizes the instantaneous CSI to allocate the total transmit power to each active subcarrier based on the KKT conditions, aiming at maximizing the average channel capacity. We analyze the formulated optimization problem and equivalently transfer it to another optimization problem for each instant. We also give two examples of adaptive OFDM IM systems with decentralized and centralized mapping scheme selections to illustrate the efficiency of the proposed power allocation strategy. Meanwhile, the general forms of the final solutions to the formulated optimization problems have also been given and their values can be provided by an iterative algorithm for a given network realization. By numerical results provided by Monte Carlo simulations, our proposed dynamic strategy can lead to a higher average network capacity than uniform power allocation strategy. By applying the proposed power allocation strategy, the efficiency of  adaptive OFDM IM can be enhanced in practice, which paves the way for its implementation in the future, especially for cell-edge communications.

\section*{Acknowledgment}
This work was supported by the SEN grant (EPSRC grant number EP/N002350/1) and the grant from China Scholarship Council  (No. 201508060323).

\bibliographystyle{IEEEtran}
\bibliography{bib}

\end{document}